\documentclass[12pt]{article}
\usepackage{jheppub,amsmath, amsthm, amssymb,slashed,url}
\usepackage{graphicx}
\usepackage{epstopdf}
\usepackage{bm}
\usepackage{kbordermatrix}
\usepackage{mathtools}

\usepackage{tikz}
\usepackage{xcolor}

\usepackage{etoolbox}
    \makeatletter
    \patchcmd{\maketitle}{\@fpheader}{}{}{}
    \makeatother

\usepackage{amsfonts}

\newcommand{\inv}[1]{\frac{1}{#1}}

\newcommand{\bra}[1]{\langle #1 |}
\newcommand{\ket}[1]{| #1 \rangle}

\newcommand{\mpi}{M_\pi}

\def\OMIT#1{{}}

\def\OMIT#1{{}}
\def\nc{{N_c}}

\def\nc{{N_c}}

\def\yo2{{f_\pi^2}}

\def\oneht{\textstyle{1\over 2} }

\newcommand{\be}{\begin{equation}}
\newcommand{\ee}{\end{equation}}
\newcommand{\bea}{\begin{eqnarray}}
\newcommand{\eea}{\end{eqnarray}}

\newcommand{\nn}{\nonumber \\}

\newcommand{\BE}{\begin{equation}}
\newcommand{\EE}{\end{equation}}
\newcommand{\BA}{\begin{eqnarray}}
\newcommand{\EA}{\end{eqnarray}}

\font\teneurm=eurm10 \font\seveneurm=eurm7 \font\fiveeurm=eurm5
\newfam\eurmfam
\textfont\eurmfam=\teneurm \scriptfont\eurmfam=\seveneurm
\scriptscriptfont\eurmfam=\fiveeurm

\font\teneusm=eusm10 \font\seveneusm=eusm7 \font\fiveeusm=eusm5
\newfam\eusmfam
\textfont\eusmfam=\teneusm \scriptfont\eusmfam=\seveneusm
\scriptscriptfont\eusmfam=\fiveeusm

\font\tencmmib=cmmib10 \skewchar\tencmmib='177
\font\sevencmmib=cmmib7 \skewchar\sevencmmib='177
\font\fivecmmib=cmmib5 \skewchar\fivecmmib='177
\newfam\cmmibfam
\textfont\cmmibfam=\tencmmib \scriptfont\cmmibfam=\sevencmmib
\scriptscriptfont\cmmibfam=\fivecmmib

\def\Pi{\varPi}

\dedicated{NT@UW-21-05, IQuS@UW-21-008}
\title{Entanglement minimization in hadronic scattering with pions}
\author{\bf Silas R.~Beane,}
\author{\bf Roland C.~Farrell,}
\author[*]{\bf{ and Mira Varma}\note[*]{Address as of 1 June 2021: Department of Physics, Yale University.}}

\affiliation{InQubator for Quantum Simulation (IQuS), Department of Physics,\\
  University of Washington, Seattle, WA 98195.}
\date{\mydate}

\abstract{Recent work~\protect\cite{Beane:2018oxh} conjectured that
  entanglement is minimized in low-energy hadronic scattering
  processes. It was shown that the minimization of the entanglement
  power (EP) of the low-energy baryon-baryon $S$-matrix implies novel
  spin-flavor symmetries that are distinct from large-$\nc$ QCD
  predictions and are confirmed by high-precision lattice QCD
  simulations. Here the conjecture of minimal entanglement is
  investigated for scattering processes involving pions and
  nucleons. The EP of the $S$-matrix is constructed for the $\pi\pi$
  and $\pi N$ systems, and the consequences of minimization of
  entanglement are discussed and compared with large-$\nc$ QCD
  expectations.}
\notoc
\begin{document} \maketitle
\newpage
\setcounter{page}{1}

\section{Introduction}
\label{sec:intro}

\noindent It is of current interest to uncover implications of quantum
entanglement for the low-energy interactions of hadrons and nuclei\footnote{For a recent review, see Ref.~\cite{Klco:2021lap}.}. As
these interactions are profitably described by effective quantum field
theory (EFT), which is an expansion of the relevant effective action
in local operators, entanglement may have subtle implications for EFT
which are difficult to identify due to its intrinsic non-locality.
Ideally entanglement properties reveal themselves as regularities in
hadronic data and, possibly, as accidental approximate symmetries. In
addition to the non-local nature of entanglement, a difficulty lies
with parsing the distinction, if any, between entanglement effects and
generic quantum correlations which account for the deviation of QCD
path integral configurations from a classical path. For instance, if
one assumes that QCD with $\nc=3$ is near the large-$\nc$
limit~\cite{tHooft:1973alw,Witten:1979kh,Witten:1979pi,Kaiser:2000gs}, then one
might expect that it would be difficult to distinguish between
large-$\nc$ expectations and some fundamental underlying principle
that minimizes entanglement independent of the value of $\nc$.  To
make this more concrete, consider two local or non-local QCD operators
${\cal O}_1$ and ${\cal O}_2$. If the vacuum expectation value of the
product of these operators
obeys the factorization rule~\cite{tHooft:1973alw,Witten:1979kh,Witten:1979pi}
\begin{equation}
\langle {\cal O}_1 {\cal O}_2 \rangle \ =\ \langle {\cal O}_1 \rangle \langle {\cal O}_2 \rangle \ +\ O(\epsilon)
	\label{eq:factorgen}
\end{equation}
where $\epsilon$ is a small number, then the variance of any operator
vanishes in the limit $\epsilon\rightarrow 0$. A theory whose
operators obey this factorization behaves like a classical
theory\footnote{Ordinarily one identifies the classical theory with
  the trivial $\hbar\rightarrow 0$ limit. However,
  Ref.~\cite{Yaffe:1981vf} has established a more general criterion
  for the classical limit.} and therefore has a small parameter
$\epsilon$ which measures quantum effects. Large-$\nc$ QCD is such a
theory, and indeed, at least for a class of QCD operators, one can
identify $\epsilon = 1/\nc$. The factorization property,
Eq.~(\ref{eq:factorgen}), is then easily deduced from Feynman diagrams
involving quarks and gluons and amounts to the dominance of
disconnected contributions in the path integral.

On the other hand, one might imagine that the factorization of
Eq.~(\ref{eq:factorgen}) arises as a property of the path integral,
rather than as a property of the local action (as in varying $\nc$ and
taking it large in QCD). It is not {\it a priori} unlikely that, at
least for a class of QCD operators, the path integral minimizes
quantum fluctuations via a mechanism that is not currently understood.
For instance, starting with QCD defined at short distances, the
procedure of sequentially integrating out short distance modes to
obtain low-energy hadronic scattering amplitudes may remove
highly-entangled states that arise from non-perturbative QCD dynamics,
leaving a low-energy EFT that is near a classical trajectory.  It is
intuitively sensible that the QCD confinement length acts as a natural
cutoff of entanglement in the low-energy EFT.  This notion can be
raised to the conjecture that QCD will minimize the entanglement in
low-energy hadronic interactions. Testing this conjecture relies on
finding hadronic systems where its consequences deviate from those
implied by large-$\nc$. And the success of the large-$\nc$ approximation
in describing the world renders this task challenging.  Evidence in
favor of this conjecture was found in Ref.~\cite{Beane:2018oxh} in a
study of baryon-baryon scattering systems (See also
Refs.~\cite{Beane:2020wjl} and \cite{Low:2021ufv}). This work relied
both on theoretical arguments and high-precision lattice QCD
simulations of baryon-baryon scattering systems with strangeness. In
this paper, the conjecture of minimal
entanglement will be investigated in both $\pi\pi$ and $\pi N$ scattering.

Finding measures of the entanglement due to interaction is both
non-trivial and non-unique.  The most fundamental object in the
scattering process is the unitary $S$-matrix.  In a scattering process
in which the two in-state particles form a product state, the
$S$-matrix will entangle the in-state particles in a manner that is
dependent on the energy of the scattering event. A useful measure of
this entanglement is the entanglement power (EP) of the
$S$-matrix~\cite{PhysRevA.63.040304,mahdavi2011cross,Beane:2018oxh}. In
the case of nucleon-nucleon ($NN$) scattering, the EP was found for all
momenta below inelastic threshold~\cite{Beane:2018oxh}. However, the
most interesting phenomenological result is at threshold, where the
vanishing EP implies the vanishing of the leading-order spin
entangling operator, which in turn implies Wigner $SU(4)$
symmetry~\cite{Wigner:1936dx,Wigner:1937zz,Wigner:1939zz}.  As this
symmetry is a consequence of large-$N_c$
QCD~\cite{Kaplan:1995yg,Kaplan:1996rk,CalleCordon:2008cz}, the
minimization of entanglement and the large-$N_c$ approximation are
found to be indistinguishable in the two-flavor case. By contrast, in
the three-flavor case, minimization of the entanglement power in
baryon-baryon scattering implies an enhanced $SU(16)$ symmetry which
is not necessarily implied by large-$N_c$ and is realized in lattice
QCD simulations~\cite{Beane:2018oxh,Low:2021ufv}. Given that
baryon-baryon scattering exhibits entanglement minimization, it is of
interest to determine whether other low-energy QCD scattering systems
exhibit this property. In investigating the EP of scattering systems
involving pions, once again a crucial difficulty is distinguishing
consequences of entanglement minimization and the large-$\nc$
limit. In the $\pi\pi$ system the implications of entanglement minimization are found to
be indistinguishable from implications of large-$N_c$. In the $\pi N$
system the implications of entanglement minimization are distinct,
however the absence of an enhanced symmetry limits the predictive
power to simple scaling laws with no smoking-gun predictions.

This paper is organized as follows. In Section~\ref{sec:pipiS}, the EP
of the $\pi\pi$ $S$-matrix is considered in detail.  After
introducting the standard definition and conventions of the $\pi\pi$
$S$-matrix, the $S$-matrix is formulated in a basis convenient for
calculation of the EP. Explicit expressions are derived for the EP of
the first few partial waves in terms of phase shifts and leading-order
expressions in chiral perturbation theory are provided.  Using the
highly-accurate Roy-equation solutions for the low-energy phase
shifts, the experimental EP for the first few partial waves are given
up to inelastic threshold. The consequences of minimizing the EP are
considered and compared to large-$\nc$ expectations.  In
Section~\ref{sec:piNS}, the same procedure is repeated for the $\pi N$
$S$-matrix. Finally, Section~\ref{sec:discuss} is a discussion of the
possible conclusions that can be drawn from the conjecture of minimal
entanglement.

\section{The $\pi\pi$ System}
\label{sec:pipiS}

\noindent There are, of course, several important differences between baryon-baryon and pion-pion scattering.
Firstly, with pions there is no notion of spin entanglement.  However,
isospin (or generally flavor) entanglement is present and can be
quantified using the EP and it is not clear that there is any
meaningful distinction between these two kinds of
entanglement. Indeed, it is straightforward to see that the ``spin''
entanglement of Ref.~\cite{Beane:2018oxh} can be reformulated as
``isospin'' entanglement with identical consequences\footnote{At the
  level of the EFT, this is simply realized via Fierz
  identities.}. This is no surprise as entanglement is fundamentally a
property of a non-product state vector whose existence relies either on an internal or a spacetime symmetry. Secondly, the crucial distinction between
baryon-baryon scattering at very low-energies and the scattering of
pions is that pion scattering at low-energies is strongly constrained
by spontaneous chiral symmetry breaking in QCD. In particular, chiral
symmetry implies that low-energy pion scattering on an arbitrary
hadronic target is weak. The weak nature of the interaction is due to
the smallness of the light-quark masses relative to a characteristic
QCD scale. This translates to a chiral suppression of the EP at
low-energies. Chiral symmetry breaking at large-$\nc$ does involve
enhanced symmetry~\cite{Kaiser:2000gs}; for $N$ flavors, the QCD
chiral symmetries and their pattern of breaking are enhanced to
$U(N)\otimes U(N)\to U(N)$, as signaled by the presence of a new
Goldstone boson, $\eta'$, whose squared mass scales as $1/\nc$.
Intuitively, the anomaly, as an intrinsically quantum phenomenon, is a
strongly entangling effect which would generally vanish as quantum
fluctuations are suppressed. However, this is not assumed as the focus of this paper is
two-body scattering which does not access the anomaly.

\subsection{$S$-matrix definition}

\noindent The $S$-matrix is defined as
\begin{equation}
	S = 1 + i T 
	\label{eq:sbasic}
\end{equation}
where unity, corresponding to no interaction, has been separated out. This then defines the $T$-matrix.
The $S$-matrix element for the scattering process $\pi^i \pi^j \to \pi^k \pi^l$ is then given by
\begin{multline}
  \bra{\pi^k(p_3) \pi^l(p_4)} S \ket{\pi^i(p_1) \pi^j(p_2)} =   \langle \pi^k(p_3) \pi^l(p_4) |  \pi^i(p_1) \pi^j(p_2) \rangle
  \\ + \bra{\pi^k(p_3) \pi^l(p_4)} iT \ket{\pi^i(p_1) \pi^j(p_2)}
\end{multline}
where $i$, $j$, $k$, and $l$ are the isospin indices of the pion states.
The projection operators onto states of definite isospin are\footnote{For a detailed construction, see Ref.~\cite{lanz2018determination}.}
\begin{eqnarray}
	P_0^{kl,ij} &=& \inv{3} \delta^{kl} \delta^{ij} \ , \\
	P_1^{kl,ij} &=& \inv{2} \left( \delta^{ki} \delta^{lj} - \delta^{li} \delta^{kj} \right) \ , \\
	P_2^{kl,ij} &=& \inv{2} \left( \delta^{ki} \delta^{lj} + \delta^{li} \delta^{kj} \right) 
								- \inv{3}\delta^{kl} \delta^{ij} \ ,
	\label{eq:pionnIP}
\end{eqnarray}
where the subscript indicates the total isospin, $I$, of the $\pi\pi$ system.
Straightforward field-theoretic manipulations then give
\begin{multline}
  \bra{\pi^k(p_3) \pi^l(p_4)} S \ket{\pi^i(p_1) \pi^j(p_2)} \\ = 
  (2\pi)^4 \delta^4(p_1+p_2-p_3-p_4)\, \frac{16 \pi}{\sigma(s)}\,\sum_{\ell=0}^\infty (2\ell+1) P_\ell(\cos \theta)\, {\bf\cal S}_\ell^{kl,ij} \ ,
\end{multline}
where the $P_\ell$ are the Legendre polynomials, and
\begin{equation}
	\sigma(s) \equiv \sqrt{1-4 \mpi^2/s} \ ,
\end{equation}
with $s=4(q^2+M_\pi^2)$ and $q$ is the center-of-mass three-momentum of the pions.
The focus here will be on the $S$-matrices of definite partial wave:
\begin{equation}
{\bf\cal S}_\ell^{kl,ij} \equiv  e^{2i\delta_\ell^0} P_0^{kl,ij}+e^{2i\delta_\ell^1} P_1^{kl,ij}+e^{2i\delta_\ell^2} P_2^{kl,ij} \ ,
\end{equation}
    \label{eq:Smatcomp}
which satisfy the unitarity constraint
\begin{equation}
{\bf\cal S}_\ell^{kl,ij} {\bf\cal S}_\ell^{*ij,mn} \ =\ \delta^{km} \delta^{ln} \ .
\end{equation}
Since the pions obey Bose statistics, the angular momentum, $\ell$, is
even for the states with $I = 0$ or $2$ and odd for states with $I = 1$.

As the initial state in the scattering process is a product state of two pions, each in the ${\bf 3}$-dimensional
($I=1$) irrep of $SU(2)$ isospin, it is convenient to work in the direct-product  matrix basis.
The pion isospin matrices are the three-by-three matrices $\hat { t}_\alpha$ which satisfy
\begin{eqnarray}
{[\,\hat { t}_\alpha\,  ,\,\hat { t}_\beta\, ]}\, =\, i\,\epsilon_{\alpha\beta\gamma} \, \hat { t}_\gamma \ .
\end{eqnarray}
In the product Hilbert space ${\cal H}_1\otimes{\cal H}_2$, the total isospin of the two-pion system is $\hat {\bm t}_1\otimes {\cal I}_3 +{\cal I}_3\otimes \hat {\bm t}_2$, where
${\cal I}_3$ is the three-by-three unit matrix, which implies
\begin{eqnarray}
  \hat {\bm t}_1 \cdot \hat {\bm t}_2 & =&  \oneht \Big\lbrack I\left(I+1\right) \;-\; 4 \Big\rbrack \hat {\bf 1}\ =\ \hat {\bf 1}
  \begin{cases}
-2,\qquad I=0 \\
-1,\qquad I=1 \\
\ \; 1,\qquad I=2
  \end{cases}
    \label{eq:basic2}
\end{eqnarray}
where
$ \hat {\bf 1} = \hat {\cal I}_3\otimes  \hat {\cal I}_3$ and
$\hat {\bm t}_1 \cdot \hat {\bm t}_2 = \sum\limits_{\alpha=1}^3 \ \hat{ t}_1^\alpha \otimes \hat{ t}_2^\alpha$.
The $9\times 9$ dimensionality of the matrix is in correspondence with the dimensionality of the $SU(2)$ isospin product representation
${\bf 3}\otimes{\bf 3}= {\bf 1}\oplus {\bf 3}\oplus {\bf 5}$. There are now three invariants and three observables; one easily finds the $S$-matrix
in the direct-product matrix basis
\begin{eqnarray}
  \hat {\bf S}_\ell   & = &
e^{2i\delta_\ell^0} \hat {\bf P}_{0}+e^{2i\delta_\ell^1} \hat {\bf P}_{1}+e^{2i\delta_\ell^2} \hat {\bf P}_{2} \ ,
\label{eq:Sdefpipi}
\end{eqnarray}
where the three $9\times 9$ projection matrices are
\begin{eqnarray}
  \hat {\bf P}_{0}& =& -\frac{1}{3}\left(\hat {\bf 1}- \left( \hat  {\bm t}_1 \cdot   \hat  {\bm t}_2 \right)^2 \right) \ , \\
    \hat {\bf P}_{1}& =& \hat {\bf 1}-\frac{1}{2}\left( \left( \hat  {\bm t}_1 \cdot   \hat  {\bm t}_2 \right)+ \left( \hat  {\bm t}_1 \cdot   \hat  {\bm t}_2 \right)^2 \right) \ , \\
        \hat {\bf P}_{2}& =& \frac{1}{3}\left(\hat {\bf 1}+ \frac{3}{2} \left( \hat  {\bm t}_1 \cdot   \hat  {\bm t}_2 \right)+ \frac{1}{2} \left( \hat  {\bm t}_1 \cdot   \hat  {\bm t}_2 \right)^2 \right) \ .
\end{eqnarray}
It is readily checked that the $S$-matrix is unitary, and using the representation $({ t}_\gamma)_{\alpha\beta}=-i\epsilon_{\alpha\beta\gamma}$, it is straightforward to establish
equivalence with the component form, Eq.~(\ref{eq:Smatcomp}).
The trace is given by $e^{i 2 \delta_\ell^0} +3 e^{i 2 \delta_\ell^1} +5 e^{i 2 \delta_\ell^2}$ which correctly
counts the isospin multiplicity, and is in correspondence with the nine eigenvalues of $\hat {\bf S}$.

\subsection{Entanglement power}
\label{sec:pipiEP}

\noindent Consider the $\ell=1$ $S$-matrix. As this system can scatter only in the $I=1$ channel, it provides
a useful example of how the $S$-matrix entangles the initial two-pion state. From Eq.~(\ref{eq:Sdefpipi}) one
finds
\begin{eqnarray}
  \hat {\bf S}_1   & = & {1\over 2}\left( 1+ e^{i 2 \delta_1^1}\right) \hat   {\bf 1}
\ +\ {1\over 2}\left( 1- e^{i 2 \delta_1^1}  \right) {\cal P}_{12}
\label{eq:Spipiell1}
\end{eqnarray}
where the SWAP operator is given by
\begin{equation}
{\cal P}_{12} =\left( \hat  {\bm t}_1 \cdot   \hat  {\bm t}_2 \right)^2 + \hat  {\bm t}_1 \cdot   \hat  {\bm t}_2 - \hat {\bf 1} \ .
\end{equation}
As the SWAP operator interchanges the pions in the initial two-pion product state, leaving another two-pion product state, it
does not entangle. Therefore, the $S$-matrix has the two obvious non-entangling solutions $\delta_1^1=0$ (no interaction) and
$\delta_1^1=\pi/2$ (at resonance). One measure of $S$-matrix entanglement would then be the (absolute value squared of the) product of the coefficients of the non-entangling
solutions:
\begin{eqnarray}
\Big\lvert\;\left( 1+ e^{i 2 \delta_1^1}\right) \left( 1- e^{i 2 \delta_1^1}\right) \Big\lvert^2\; \sim \sin^2\left(2 \delta_1^1\right) \ .
\label{eq:EPnaive}
\end{eqnarray}

A state-independent measure of the entanglement generated by the
action of the $S$-matrix on the initial product state of two free
particles is the
EP~\cite{PhysRevA.63.040304,mahdavi2011cross,Beane:2018oxh}.  In order
to compute the EP an arbitrary initial product state should be
expressed in a general way that allows averaging over a given
probability distribution folded with the initial state. Recall that in
the $NN$ case, there are two spin states (a qubit) for each nucleon and
therefore the most general initial nucleon state involves two complex
parameters or four real parameters. Normalization gets rid of one
parameter and there is an overall irrelevant phase which finally
leaves two real parameters which parameterize the ${\bf CP}^1$
manifold, also known as the 2-sphere ${\bf S}^2$, or the Bloch sphere.
Now in the isospin-one case we have three isospin states (a qutrit)
which involves three complex parameters. Again normalization and
removal of the overall phase reduce this to four real parameters which
parameterize the ${\bf CP}^2$
manifold~\cite{Brody_2001,Bengtsson:2001yd,bengtsson_zyczkowski_2006}. Since
the $\pi\pi$ initial state is the product of two isospin-one states,
there will be eight parameters to integrate over to get the EP.

There are now two qutrits in the initial state, which live in the Hilbert spaces ${\cal H}_{1,2}$, each
spanned by the states $\lbrace |\,{\bf -1}_i \, \rangle , |\,{\bf 0}_i
\, \rangle , |\,{\bf 1}_i \, \rangle \rbrace$ with $i=1,2$.  It is of interest
to determine the EP of a given $S$-matrix operator, which is a measure of the entanglement of the scattered state averaged
over the ${\bf CP}^2$ manifolds on which the qutrits live.  Consider
an arbitrary initial product state of the qutrits
\begin{eqnarray}
|\,\Psi \, \rangle  \ =\  U\left(\alpha_1,\beta_1,\mu_1,\nu_1\right) | \, \rangle_1 \otimes U\left(\alpha_2,\beta_2,\mu_2,\nu_2\right) | \, \rangle_2 
\end{eqnarray}
with
\begin{eqnarray}
\hspace{-0.34in}U\left(\alpha_i,\beta_i,\mu_i,\nu_i\right) | \, \rangle_i  =
\cos\beta_i \sin\alpha_i|\,{\bf -1} \, \rangle_i  +  e^{i\mu_i}\sin\beta_i \sin\alpha_i|\,{\bf 0} \, \rangle_i  + e^{i\nu_i}\cos\alpha_i |\,{\bf 1} \, \rangle_i\, ,
\end{eqnarray}
where $0 \leq {\mu_i,\nu_i} < 2{\pi}$ and $0 \leq {\alpha_i,\beta_i} \leq {\pi}/2$.
The geometry of ${\bf CP}^2$ is described by the Fubini-Study (FS) line element~\cite{Brody_2001,Bengtsson:2001yd,bengtsson_zyczkowski_2006}
\begin{eqnarray} 
\begin{split}
ds_{\scriptstyle FS}^2 = \; &d\alpha^2+ \sin ^2(\alpha ) d\beta^2 +\left ( \sin ^2(\alpha ) \sin ^2(\beta ) - \sin ^4(\alpha ) \sin ^4(\beta ) \right ) d\mu^2 + \\
& \sin ^2(\alpha ) \cos ^2(\alpha ) d\nu^2 -2 \sin ^2(\alpha ) \cos ^2(\alpha ) \sin ^2(\beta ) d\mu  d\nu \ .
\end{split}
\label{FSmetric}
\end{eqnarray}
Of special interest here is the differential volume element which in these coordinates is
\begin{eqnarray}
  dV_{\scriptstyle FS} &=& \sqrt{det\,g_{\scriptstyle FS}}\, d\alpha\, d\beta\, d\mu\, d\nu  \nonumber \\
  &=&\cos\alpha\cos\beta\sin^3\alpha\sin\beta\, d\alpha\, d\beta\, d\mu\, d\nu \  
\end{eqnarray}
and the volume of the ${\bf CP}^2$ manifold is found to be,
\begin{eqnarray}
\int  dV_{\scriptstyle FS} & =& \frac{\pi^2}{2} \ .
\end{eqnarray}

The final state of the scattering process is obtained by acting with the unitary $S$-matrix of definite angular momentum
on the arbitrary initial product state:
\begin{eqnarray}
|\,\bar \Psi \, \rangle  \ =\  \hat {\bf S}_\ell |\, \Psi \, \rangle  \ .
\end{eqnarray}
The associated density matrix is
\begin{eqnarray}
\rho_{1,2}   \ =\  |\, \bar \Psi \, \rangle  \langle \, \bar \Psi  | \, ,
\end{eqnarray}
and tracing over the states in ${\cal H}_{2}$ gives the reduced density matrix
\begin{eqnarray}
\rho_{1}   \ =\  {\rm Tr}_2 \big\lbrack \rho_{1,2}  \big\rbrack .
\end{eqnarray}
The linear entropy of the $S$-matrix is then defined as\footnote{Note that this is related to the (exponential of the) second R\'enyi entropy.}
\begin{eqnarray}
E_{\hat {\bf S}_\ell}  \ =\  1 \ -\ {\rm Tr}_1 \big\lbrack \left( \rho_{1} \right)^2 \big\rbrack .
\end{eqnarray}
Finally, the linear entropy is integrated over the initial ${\bf CP}^2$ manifolds to form the average, and the entanglement power is
\begin{eqnarray}
  {\mathcal E}({\hat {\bf S}_\ell}) \ =\ \left(\frac{2}{\pi^2}\right)^2 \left(\prod_{i=1}^2\int dV^i_{\scriptstyle FS}\right) {\cal P} E_{\hat {\bf S}_\ell} 
\end{eqnarray}
where ${\cal P}$ is a probability distribution which here will be taken to be unity.
Evaluating this expression using Eq.~(\ref{eq:Sdefpipi}) yields the s-wave $\pi\pi$ EP:
\begin{eqnarray}
 {\mathcal E}({\hat {\bf S}_0}) & =& \frac{1}{648}\left( 156 - 6 \cos[4 \delta_0^0 ] - 65 \cos[2 (\delta_0^0 - \delta_0^2)] \right. \nn 
  &&\qquad\qquad\qquad\left.\hspace{-0.48in}- 10 \cos[4 (\delta_0^0 - \delta_0^2)] - 60 \cos[4 \delta_0^2] - 
 15 \cos[2 (\delta_0^0  + \delta_0^2)] \right)\ ,
\label{eq:EPswave}
\end{eqnarray}
and the p-wave $\pi\pi$ EP:
\begin{eqnarray}
  {\mathcal E}({\hat {\bf S}_1})  = \frac{1}{4}\sin^2\left(2 \delta_1^1\right) \ .
\label{eq:EPpwave}
\end{eqnarray}
Notice that this matches the intuitive construction which led to Eq.~(\ref{eq:EPnaive}).
The EPs have the non-entangling solutions:
\begin{eqnarray}
\delta_0^0 & =& \delta_0^2 \ =\ 0, \frac{\pi}{2} \ , \\
\delta_1^1 & =& 0, \frac{\pi}{2} \ .
\label{eq:noEPsols2bose}
\end{eqnarray}
Therefore, in the s-wave, entanglement minimization implies that {\it
  both} isospins are either non-interacting or at resonance, while in
the p-wave, entanglement minimization implies that the $I=1$ channel
is either non-interacting or at resonance. As no $I=2$ resonances are
observed in nature (and their coupling to pions is suppressed in
large-$\nc$ QCD~\cite{Weinberg:2013cfa}), the s-wave EP has a single minimum corresponding to
no interaction. By contrast, the $I=1$ channel will exhibit minima of
both types. It is worth considering the EP of a simple resonance model.
Consider the unitary $S$-matrix:
\begin{eqnarray}
  {\hat {\bf S}_1} \ =\ \frac{s-m_1^2 - i m_1 \Gamma_1}{s-m_1^2 + i m_1 \Gamma_1} \ ,
\end{eqnarray}
where $m_1$ ($\Gamma_1$) are the mass (width) of the resonance. The EP is
\begin{eqnarray}
  &&{\mathcal E}({\hat {\bf S}_1})  = \left( \frac{m_1 \Gamma_1 \left(s-m_1^2\right)}{\left(m_1 \Gamma_1\right)^2+ \left(s-m_1^2\right)^2}\right)^2 \ ,
\end{eqnarray}
which vanishes on resonance at $s=m_1^2$ and has maxima at $s=m_1(m_1\pm \Gamma_1)$. It is clear that the minimum corresponds to
$\hat {\bf S}\propto {\cal P}_{12}$. As the $\rho$-resonance dominates the $I=1$ channel at energies below $1~{\rm GeV}$, the EP in nature
will be approximately of this form.

\begin{figure}[!h]
\centering
\includegraphics[width = 0.81\textwidth]{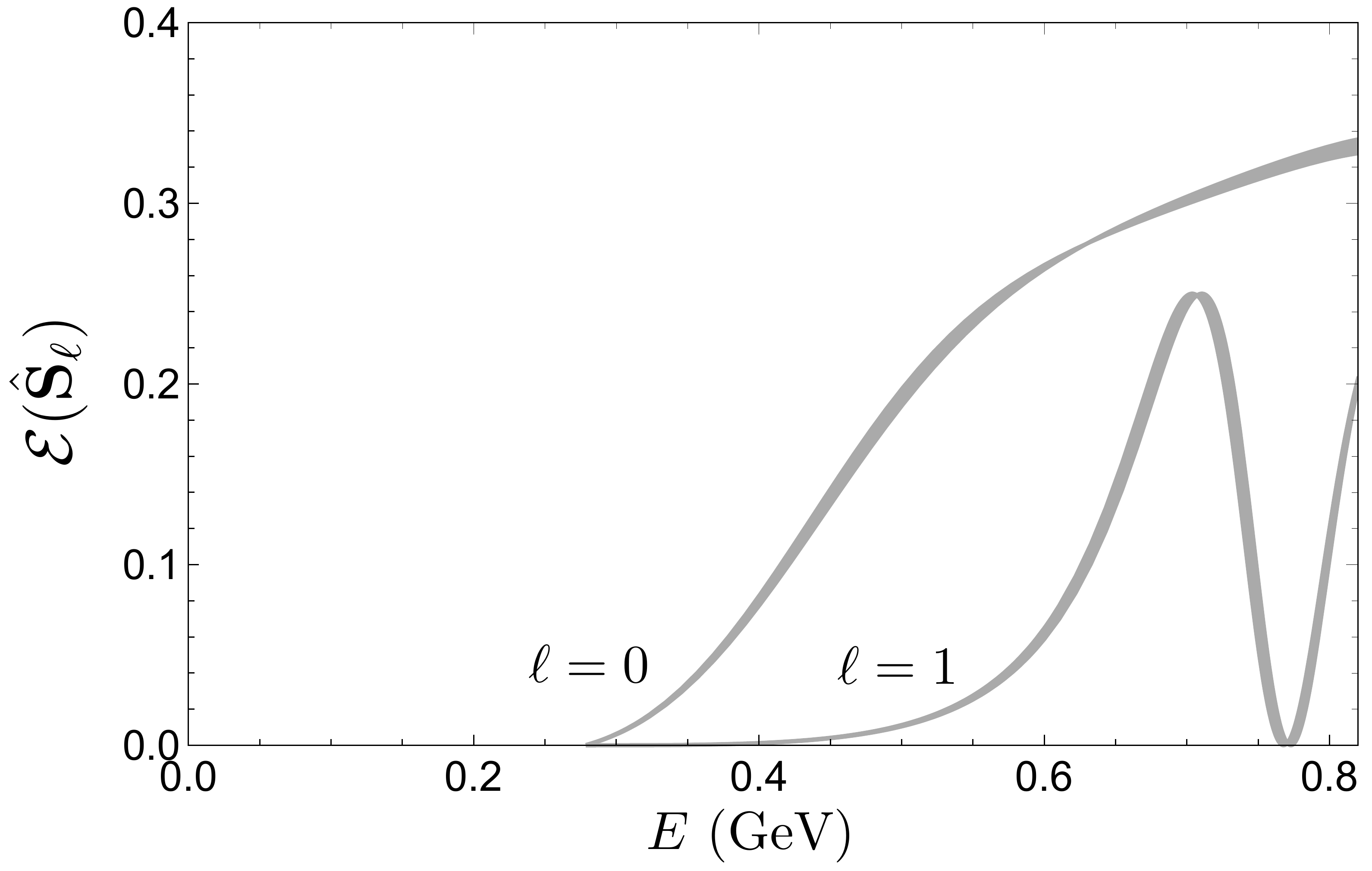}
\caption{Entanglement power of the $\pi\pi$ $S$-matrix for $\ell=0,1$ taken from Roy equation determinations (the bands represent an estimate of the uncertainties~\cite{Ananthanarayan:2000ht,Colangelo:2001df})
  of the $\pi\pi$ phase shifts.}
    \label{fig:EPpipiROY}
\end{figure}

The $\pi\pi$ phase shifts are the most accurately known of all hadronic $S$-matrices as the Roy equation constraints~\cite{Roy:1971tc}
come very close to a complete determination of the phase shifts~\cite{Ananthanarayan:2000ht,Colangelo:2001df}.
In Fig.~(\ref{fig:EPpipiROY}) the EPs for the first few partial waves are plotted using the Roy equation determinations of the $S$-matrix.

\subsection{Chiral perturbation theory}

\noindent Near threshold, the phase shift can be expressed in the effective range expansion as
\begin{equation}
	\delta_\ell^I(s) = \oneht \sin^{-1} \lbrace 2\sigma (s) q^{2\ell}\left( a^I_\ell \;+\; \mathcal{O}(q^2) \right)\rbrace \ ,
\end{equation}
where the scattering lengths, $a^I_\ell$, relevant to s-wave and p-wave scattering, are given at leading order in chiral perturbation theory by~\cite{Weinberg:1978kz,Gasser:1983yg}
\begin{equation}
a_0^0\ =\ \frac{7M_\pi^2}{32\pi F_\pi^2}  \ \ , \ \   a_0^2\ =\ -\frac{M_\pi^2}{16\pi F_\pi^2} \ \ , \ \ a_1^1\ =\ \frac{1}{24\pi F_\pi^2} \ ,
\end{equation}
where $F_\pi=93~{\rm MeV}$ is the pion decay constant. Near threshold the s-wave and p-wave EPs are given by
\begin{eqnarray}
&&{\mathcal E}({\hat {\bf S}_0}) \ =\ \frac{1}{9 M_\pi^2}\big\lbrack 4(a^0_0)^2 -5 (a^0_0 a^2_0)+10 (a^2_0)^2 \big\rbrack\; q^2 \ +\ \mathcal{O}(q^4)\ , \nn
&&{\mathcal E}({\hat {\bf S}_1}) \ =\ \frac{1}{M_\pi^2} (a^1_1)^2 \; q^6 \ +\ \mathcal{O}(q^{8})  \ .
\end{eqnarray}
As $a_0^0$ ($a_0^2$ ) is positive (negative) definite, the EP is trivially minimized with vanishing scattering lengths.
This then implies a bookkeeping where $F_\pi=\mathcal{O}(\epsilon^{-n})$ where $n$ is a positive number. 
Hence, in the limit of vanishing entanglement, the pions are non-interacting, and the dominant interaction is from tree diagrams; i.e. loops are suppressed by inverse powers
of $F_\pi$. In the large-$\nc$ limit, one finds  $\epsilon=1/N_c$ and $n=1/2$~\cite{tHooft:1973alw,Witten:1979kh,Witten:1979pi}.
Evidently the implications of vanishing entanglement for the $\pi\pi$ $S$-matrix are indistinguishable from large-$\nc$ expectations\footnote{We also studied the effect of explicit
chiral symmetry breaking on the entanglement power by varying the coefficients of operators with
insertions of the quark mass matrix in the effective action. No evidence of a connection between chiral symmetry breaking and the entanglement power was found. This aligns with large-$\nc$
expectations as the meson masses are independent of $\nc$. For an example of a relationship between
entanglement and chiral symmetry breaking see~\cite{Beane:2019loz}.}.

\section{The $\pi N$ System}
\label{sec:piNS}

\noindent As baryons are formed from $\nc$ quarks, the baryon masses
and axial matrix elements grow with $\nc$.  The unitarity of the
$S$-matrix then places powerful constraints on baryon properties via
large-$\nc$ consistency
conditions~\cite{Dashen:1993as,Dashen:1993ac,Dashen:1993jt,Dashen:1994qi}.
At leading order in the large-$\nc$ expansion this yields predictions
that are equivalent in the two (three) flavor case to $SU(4)$
($SU(6)$) spin-flavor symmetry which place the ground-state baryon
spin states in the ${\bf 20}$ (${\bf 56}$) dimensional irrep together
with the delta (baryon decuplet). Therefore, the large-$\nc$ limit not
only predicts an enhanced symmetry but also alters the definition of
a baryon in a fundamental way. Moreover, any sensible effective
theory of $\pi N$ scattering in the large-$\nc$ limit must include the
delta resonance as an explicit degree of freedom. In what follows, the
consequences of entanglement minimization of the low-energy $S$-matrix
are considered for $\nc=3$ QCD.

\subsection{$S$-matrix definition}
The $S$-matrix element for the scattering process, $\pi^a(q_1) N(p_1) \to \pi^b(q_2) N(p_2)$, is given by
\begin{multline}
  \bra{\pi^b(q_2) N(p_2)} S \ket{\pi^a(q_1) N(p_1)} =   \langle \pi^b(q_2) N(p_2) |  \pi^a(q_1) N(p_1) \rangle
  \\ + \bra{\pi^b(q_2) N(p_2)} iT \ket{\pi^a(q_1) N(p_1)} ,
\end{multline}
where $a$ and $b$ label the isospin of the pion.
The $T$ matrix element in the center-of-mass system (cms) for the process may be parameterized as~\cite{Fettes_1998}
\begin{equation}
    \begin{split}
        T^{ba}_{\pi N} = \left( {E+m\over 2m} \right ) \bigg \{ \delta^{ba} &\left [ g^+(\omega,t) + i \vec{\sigma} \cdot (\vec{q_2} \times \vec{q_1}) h^+(\omega,t) \right ] \\
         + i\epsilon^{abc} \tau^c &\left [ g^-(\omega,t) + i \vec{\sigma} \cdot (\vec{q_2} \times \vec{q_1}) h^-(\omega,t) \right ] \bigg \}
    \end{split}
\end{equation}
where $E$ is the nucleon energy, $\omega$ is the pion energy, $m$ is
the nucleon mass and $t = (q_1 - q_2)^2$ is the square of the momentum
transfer. The $\sigma$($\tau$) matrices act on the spin(isospin) of
the incoming nucleon. This decomposition reduces the scattering
problem to calculating $g^{\pm}$, the isoscalar/isovector
non-spin-flip amplitude and $h^{\pm}$, the isoscalar/isovector
spin-flip amplitude. The amplitude can be further projected onto
partial waves by integrating against $P_{\ell}$, the relevant Legendre
polynomial:
\begin{equation}
    f_{\ell \pm}^{\pm}(s) = {E + m \over 16\pi \sqrt{s}} \int_{-1}^{+1} dz \left [ g^{\pm} P_{\ell}(z) + {\vec{q}}^{\, 2} h^{\pm} \big ( P_{\ell \pm 1}(z) -z P_{\ell}(z)\big ) \right ]\ .
\end{equation}
Here $z=\cos{\theta}$ is the cosine of the scattering angle, $s$ is
the cms energy squared and $\vec{q}^{\, 2} = \vec{q_1}^2 =
\vec{q_2}^2$. The subscript $\pm$ on the partial wave amplitude
indicates the total angular momentum $J = \ell \pm s$. The amplitudes
in the total isospin $I=\frac{1}{2}$ and $I=\frac{3}{2}$ can be
recovered via the identification:
\begin{equation}
    f_{\ell \pm}^{\frac{1}{2}} = f_{\ell \pm}^+ + 2 f_{\ell \pm}^- \ \ , \ \ f_{\ell \pm}^{\frac{3}{2}} = f_{\ell \pm}^+ - f_{\ell \pm}^- \ .
\end{equation}
Below inelastic threshold the scattering amplitude is related to a unitary $S$-matrix by
\begin{equation}
    S_{\ell \pm}^I(s) = 1 + 2 i \lvert \, \vec{q} \, \rvert f_{\ell \pm}^I(s) \ \ , \ \ S_{\ell \pm}^I(s) S_{\ell \pm}^I(s)^{\dagger} = 1 
\end{equation}
and the $S$-matrix can be parameterized in terms of phase shifts,
\begin{equation}
    S_{\ell \pm}^I(s) = e^{2 i \delta_{\ell \pm}^I(s)} \ .
    \label{eq:pinsmatrix}
\end{equation}
For a more detailed derivation of the $\pi N$ $S$-matrix
see~\cite{Fettes_1998,Yao_2016,Moj_i__1998,Scherer2012}. Scattering in
a given partial wave and total angular momentum channel leads to a
$S$-matrix which acts on the product Hilbert space of
the nucleon and pion isospin, $\mathcal{H}_{\pi} \otimes
\mathcal{H}_{N}$. The $S$-matrix can then be written in terms of total
isospin projection operators
\begin{equation}
    \hat{{\bf S}}_{\ell \pm} = e^{2 i \delta_{\ell \pm}^{3/2}} \hat{{\bf P}}_{3/2} + e^{2 i \delta_{\ell \pm}^{1/2}} \hat{{\bf P}}_{1/2}
\end{equation}
where the $6 \times 6$ projection matrices are
\begin{equation}
    \begin{split}
        &\hat{{\bf P}}_{3/2} = \frac{2}{3} \left (\hat{{\bf 1}} + \hat{{\bf t}}_{\pi} \cdot \hat{{\bf t}}_N \right ) \ , \\
        &\hat{{\bf P}}_{1/2} = \frac{1}{3} \left (\hat{{\bf 1}} - 2(\hat{{\bf t}}_{\pi} \cdot \hat{{\bf t}}_N) \right )\ .
    \end{split}
\end{equation}
The operators $\hat{{\bf t}}_N$ and $\hat{{\bf t}}_{\pi}$ are in the
$2$ and $3$ dimensional representations of $SU(2)$ isospin
respectively and $\hat {\bm t}_{\pi} \cdot \hat {\bm t}_N =
\sum\limits_{\alpha=1}^3 \ \hat{ t}_{\pi}^\alpha \otimes \hat{ t}_N^\alpha$.

\subsection{Entanglement power}
\begin{figure}
    \centering
    \includegraphics[width = 1 \textwidth]{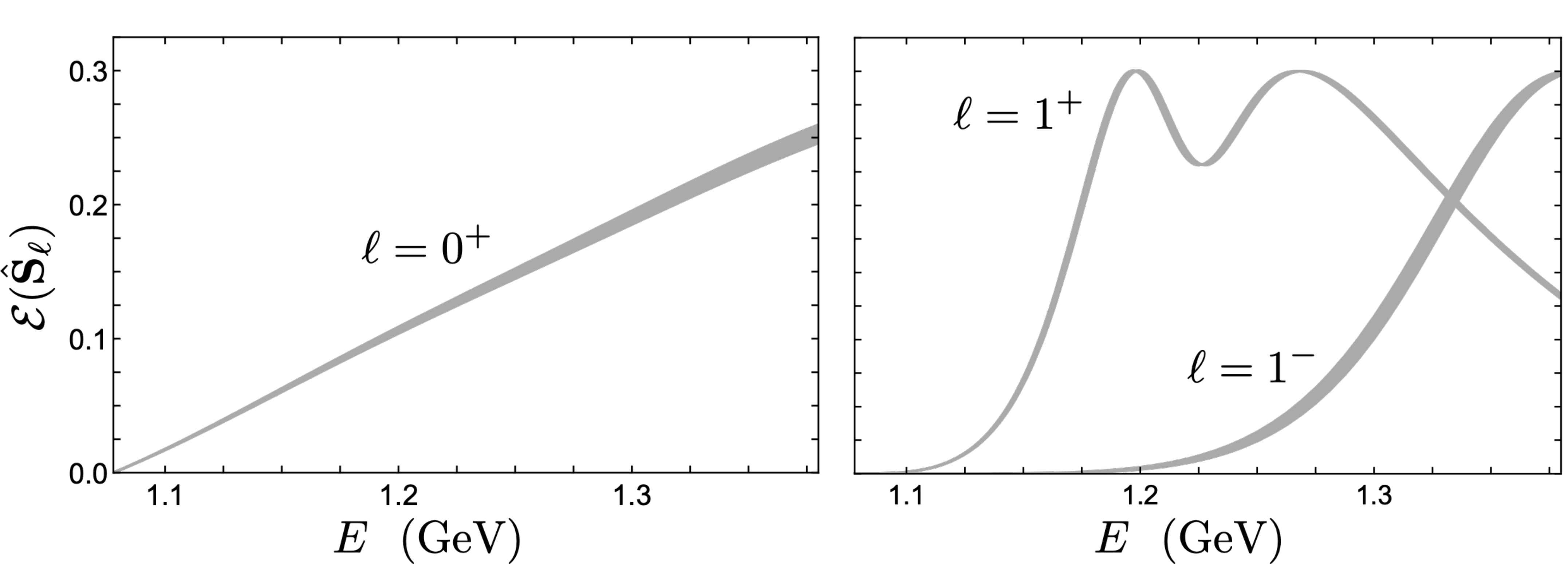}
    \caption{Entanglement power of the $\pi N$ $S$-matrix for $\ell = 0 , 1$ taken from Roy-Steiner equation determinations (the bands represent an estimate of the uncertainties~\cite{Hoferichter_2016}) of the $\pi N$ phase shifts.}
     \label{fig:piNEP}
\end{figure}

The entanglement power of the $\pi N$ $S$-matrix can be computed in a
similar manner as for the $\pi \pi$ EP. The incoming separable state
now maps to a point on the product manifold, ${\bf CP}^2 \times {\bf
  S}^2$. The construction of the reduced density matrix follows the
same steps as in section \ref{sec:pipiEP} and the entanglement power
is found to be,
\begin{equation}
\begin{split}
{\mathcal E}({\hat {\bf S}}_{\ell \pm}) &= \left ( \frac{2}{\pi^2}\frac{1}{4 \pi} \right ) \left (\int dV_{FS} d \Omega \right ) \mathcal{P} E_{\hat {\bf S}_{\ell \pm}} \\
&=
\frac{8}{243}  \bigg[ 17 + 10 \cos \left ( 2 \big ( \delta_{\ell \pm}^{{3/2}} - \delta_{\ell \pm}^{{1 / 2}}\big ) \right ) \bigg] \sin ^2\left (\delta_{\ell \pm}^{{3 / 2}} - \delta_{\ell \pm}^{{1 / 2}} \right )
\end{split}
\end{equation}
where $\mathcal{P}$ has been taken to be $1$.  Note that the two
particles are now distinguishable and so scattering in each partial
wave is no longer constrained by Bose/Fermi statistics. It follows
that the $S$-matrix is only non-entangling when it is proportional to
the identity which occurs when,
\begin{equation}
    \delta_{\ell \pm}^{{3 / 2}} = \delta_{\ell \pm}^{{1 / 2}} \ .
\end{equation}
Notice that the EP allows for interesting local minima when the difference in $I=3/2$ and $I=1/2$ phase shifts is $\pi/2$.
The $\pi N$ phase shifts are determined very accurately by the
Roy-Steiner equations up to a center-of-mass energy of 1.38
GeV~\cite{Hoferichter_2016} and the entanglement power for the first
couple partial waves is shown in Fig.~(\ref{fig:piNEP}).
There is a local minimum near the delta resonance position in the p-wave due to the rapid change of the
$I=3/2$ phase shift.

\subsection{Chiral perturbation theory}
Near threshold the phase shifts can be determined by the scattering lengths through the effective range expansion,
\begin{equation}
   \delta_{\ell \pm}^I = \cot^{-1}{\left \{ {1\over  \lvert \, \vec{q} \, \rvert ^{2 \ell +1} }\left ( {1 \over a_{\ell \pm}^I}  + \mathcal{O}(\vec{q}^{\, 2}) \right ) \right \} }\ .
\end{equation}
This leads to the threshold form of the entanglement power,
\begin{equation}
{\mathcal E}({\hat {\bf S}}_{\ell \pm}) = \frac{8}{9} \left (a^{1\over2}_{\ell \pm}- a^{3\over2}_{\ell \pm}\right )^2 \vec{q}^{\, 2+4l}
\end{equation}
which can only vanish if $a^{1\over2}_{\ell \pm} = a^{3\over2}_{\ell \pm}$.
The scattering lengths at leading order in heavy-baryon chiral perturbation theory, including the delta, are given by~\cite{Fettes_1998, Fettes_2001},
\begin{equation}
    \begin{split}
    &a^{1\over2}_{0+} = \frac{2 M_{\pi} m}{8 \pi (m + {M_{\pi}}) F_{\pi}^2} \ \ , \ \ a^{3\over2}_{0+} = \frac{- M_{\pi} m}{8 \pi (m + {M_{\pi}}) F_{\pi}^2} \\
    &a^{1\over2}_{1-} = -\frac{m \left(9  g_A^2 \Delta +9 g_A^2 M_{\pi}-8 g_{\pi N \Delta}^2 M_{\pi}\right)}{54 \pi  F_{\pi}^2 M_{\pi} (\Delta+M_{\pi}) (m+M_{\pi})} \ \ , \ \ 
    a^{3\over2}_{1-} =-\frac{m \left(9  g_A^2 \Delta +9 g_A^2 M_{\pi}-8 g_{\pi N \Delta}^2 M_{\pi}\right)}{216 \pi  F_{\pi}^2 M_{\pi} (\Delta+M_{\pi}) (m+M_{\pi})} \\
    &a^{1\over2}_{1+} = \frac{m \left(-3  g_A^2 \Delta +3 g_A^2 M_{\pi}+8 g_{\pi N \Delta}^2 M_{\pi}\right)}{72 \pi  F_{\pi}^2 M_{\pi} (\Delta-M_{\pi}) (m+M_{\pi})} \ \ , \ \
    a^{3\over2}_{1+} = \frac{m \left(-3 g_A^2 \Delta^2 -2  g_{\pi N \Delta}^2 M_{\pi} \Delta +3 g_A^2 M_{\pi}^2\right)}{36 \pi  F_{\pi}^2 M_{\pi} \left(M_{\pi}^2-\Delta^2\right) (m+M_{\pi})}
    \end{split}
\end{equation}
where $\Delta = m_{\Delta} - m_N$ is the delta-nucleon mass splitting. The corresponding EPs near threshold are,
\begin{equation}
\begin{split}
&{\mathcal E}({\hat {\bf S}}_{0+}) =  \frac{m^2 M_{\pi}^2}{8 \pi ^2 F_{\pi}^4 (m+M_{\pi})^2}\vec{q}^{\, 2} \\
&{\mathcal E}({\hat {\bf S}}_{1-}) = \frac{m^2  \left(9  g_A^2 \Delta +9 g_A^2 M_{\pi}-8 g_{\pi N \Delta}^2 M_{\pi}\right)^2}{5832 \pi ^2 f_{\pi}^4 M_{\pi}^2 (\Delta+M_{\pi})^2 (m+M_{\pi})^2} \vec{q}^{\, 6}\\
&{\mathcal E}({\hat {\bf S}}_{1+}) = \frac{m^2  \left(-9  g_A^2 \Delta^2 +4  g_{\pi N \Delta}^2 \Delta M_{\pi}+ \left(9 g_A^2+8 g_{\pi N \Delta}^2\right) M_{\pi}^2 \right)^2}{5832 \pi ^2 f_{\pi}^4 M_{\pi}^2 \left(M_{\pi}^2-\Delta^2\right)^2 (m+M_{\pi})^2} \vec{q}^{\, 6}\ .
\end{split}
\end{equation}
Once again the only non-entangling solution consistent with chiral
symmetry is no interaction, with the same scaling of $F_\pi$ as found in $\pi\pi$ scattering.
Unlike the large-$\nc$ limit, there is no reason to expect an enhancement of the axial couplings,
which in that case gives rise to the contracted spin-flavor symmetries~\cite{Dashen:1993as}.

\section{Discussion}
\label{sec:discuss}

\noindent In QCD the number of colors, $\nc$, is a parameter that
appears in the action and in some sense acts as a knob that dials the
amount of quantum correlation in the hadronic $S$-matrix.  The
simplifications, counting rules and enhanced symmetries implied by the
large-$\nc$ approximation have proved highly successful in explaining
regularity in the hadronic spectrum. Recent work in
Ref.~\cite{Beane:2018oxh} has conjectured that, independent of the
value of $\nc$, quantum entanglement is minimized in hadronic
$S$-matrices. Verifying this conjecture relies on finding consequences
of the conjecture that are distinct from large-$\nc$ predictions, and
indeed this has been found to be the case in baryon-baryon scattering.  In
particular, minimization of entanglement near threshold leads to
enhanced symmetry that is verified by lattice QCD simulations.  Here
this conjecture has been considered for $\pi\pi$ and $\pi N$
scattering.  As shown long ago by Weinberg, the scattering of soft
pions off any target is completely determined by chiral
symmetry~\cite{PhysRevLett.17.616} and is weak at low energies.  Here
it has been found that the only $\pi \pi$ or $\pi N$ $S$-matrix,
consistent with the low energy theorems, that does not entangle
isospin is the identity i.e. no scattering. In the context of chiral
perturbation theory this corresponds to $F_{\pi}$ being large when
entanglement is minimized, consistent with large $\nc$ scaling. Unlike
in the large $\nc$ limit, entanglement minimization of the $S$-matrix
says nothing about the scaling of the baryon masses and axial couplings and therefore
implies no new symmetries in the $\pi N$ sector.  Because of the
weakness of pion processes implied by chiral symmetry, it may be the
case that only systems without external Goldstone bosons (like $NN$)
give non-trivial constraints from entanglement minimization.
Considering general meson-nucleon scattering, it is clear that
scalar-isoscalar mesons have no spin or isospin to entangle.
Insofar as resonance saturation is effective, entanglement minimization would then predict the contribution to baryon-baryon scattering from
the exchange of non scalar-isoscalar resonances to sum together to give an equal
contribution to all spin-isospin channels~\cite{Epelbaum_2002}. This would then naturally lead to the $SU(16)$ symmetry seen in the three flavor baryon
sector~\cite{Beane:2018oxh}.

Techniques which make use of entanglement minimization to select out
physically-relevant states and operators from an exponentially large
space have a long history. For instance, tensor methods and DMRG
crucially rely on the fact that ground states of reasonable
Hamiltonians often exhibit much less entanglement than a typical
state~\cite{Eisert_2010}. In nuclear physics it has recently been
shown that entanglement is a useful guiding principle when
constructing many-body wave-functions of atomic
nuclei~\cite{Robin:2020aeh}.  The use of entanglement minimization to
constrain hadronic $S$-matrices in other contexts has also been
investigated recently. The authors
of~\cite{Bose:2020cod,10.21468/SciPostPhys.9.5.081} have considered
entanglement minimization as an ingredient in an effort to revive the
$S$-matrix bootstrap program. When applied to the $\pi \pi$ $S$-matrix
a correspondence is found between minima of entanglement and linear
Regge trajectories. This is an intriguing prospect and may be related
to the observation made here that, at least in p-wave $\pi \pi$
scattering, the entanglement power has a zero at resonance. Outside of
hadronic physics, it was shown recently that the minimization of spin
entanglement in scattering due to the exchange of gravitons picks out
parameters which correspond to minimally coupled
gravity~\cite{Aoude_2020}.

An interesting and directly related line of inquiry is the connection
between entanglement and renormalization group (RG) flow. As a zeroth
order observation, macroscopic objects are distinguishable from their
surroundings despite being coupled to the environment. Therefore, in some sense classical objects behave like coherent quantum states whose entropy
does not increase when they interact with an open quantum
system~\cite{PhysRevLett.70.1187}. This “motivates” the idea that at
large scales a fixed point of the entanglement entropy is reached. From an RG
point of view this may be manifest in the entanglement structure
between different momentum modes of fields. Work has been done on
computing the momentum space entanglement in both scattering events
and between regions of the ground state of a quantum field
theory~\cite{Peschanski_2016,Seki_2015,Balasubramanian_2012}. It is
speculated that the RG flow of parameters in an effective action is
driven by entanglement between the IR and UV. Along a similar vein, recent
work has employed numerical methods to study the ground state entanglement
structure between disjoint regions of massless free scalar field theory~\cite{Klco_2021,klco2021entanglement}. In tension with the tenets of EFT, it
was found that long distance entanglement gets most of its support from short
distance field modes. With this in mind it may be possible that EFT, with its insensitivity to
physics at the cutoff, is not the best framework with which to study entanglement.
There is clearly much to explore on the relationship
between entanglement, RG flow and EFT.

\section*{Acknowledgments}

\noindent We would like to thank Natalie Klco and Martin J.~Savage for valuable
discussions. This work was supported by the U.~S.~Department of Energy
grants {\bf DE-FG02-97ER-41014} (UW Nuclear Theory) and {\bf DE-SC0020970}
(InQubator for Quantum Simulation).

\bibliographystyle{JHEP}
\bibliography{bibi}

\end{document}